\documentclass[a4paper,fleqn]{cas-sc}
\usepackage[authoryear]{natbib}

\def\tsc#1{\csdef{#1}{\textsc{\lowercase{#1}}\xspace}}
\tsc{WGM}
\tsc{QE}
\tsc{EP}
\tsc{PMS}
\tsc{BEC}
\tsc{DE}

\begin{document}
\title [mode=title]{The role of ocean circulation in driving hemispheric symmetry breaking of the ice shell of Enceladus}
\let\WriteBookmarks\relax
\def\floatpagepagefraction{1}
\def\textpagefraction{.001}
\shortauthors{Kang et~al.}
\shorttitle{Ocean Circulation on Icy Moons}
\author[1]{Wanying Kang}
\ead{wanying@mit.edu}
\address[1]{Earth, Atmospheric, and Planetary Sciences, Massachusetts Institute of Technology, Cambridge, MA 02139 US}
\credit{Conceptualization, Methodology, Modeling, Data analysis, Visualization, Writing - original draft}

\author[1]{ Suyash Bire}
\ead{bire@mit.edu}
\credit{Methodology}

\author[1]{ John Marshall}
\ead{jmarsh@mit.edu}
\credit{Conceptualization, Writing - review \& editing}

\begin{abstract}
  The ice shell of Enceladus exhibits strong asymmetry between its hemispheres, with all known geysers concentrated over the south pole, even though its orbital configuration is almost perfectly symmetric. By exploring ocean circulation across a range of ocean salinities and core/shell heating partitions, we study the role of ice-ocean interaction in hemispheric symmetry breaking. We find that: (i) asymmetry is enhanced by cross-equatorial ocean heat transport when the ice shell is the major heat source and vice versa, (ii) the magnitude of ocean heat transport is comparable to the global heat production, significantly affecting the ice shell evolution and equilibrium state and (iii) more than one equilibrium state can exist due to a positive feedback between melting and ocean circulation. 
\end{abstract}

\begin{highlights}
\item Ocean circulation can transport heat across the equator supporting observed inter-hemispheric asymmetries in ice shell thickness.
\item Asymmetry in ice shell thickness is enhanced by ocean circulation if tidal heating occurs primarily in the ice shell rather than the core.
\item The coupled ocean-ice system may have multiple equilibria.
\end{highlights}


\maketitle

\section{Introduction}
  Beneath the ice shell encasing Enceladus, a small icy moon of Saturn, a global ocean of liquid water \citep{Postberg-Kempf-Schmidt-et-al-2009:sodium, Thomas-Tajeddine-Tiscareno-et-al-2016:enceladus} ejects geyser-like sprays into space through fissures concentrated near the south pole \citep{Hansen-Esposito-Stewart-et-al-2006:enceladus, Howett-Spencer-Pearl-et-al-2011:high, Spencer-Nimmo-2013:enceladus, Waite-Glein-Perryman-et-al-2017:cassini}, making it one of the places with the highest potential of finding extra-terrestrial life. However, it remains a puzzle as to why all geysers concentrate near the south pole \citep{Howett-Spencer-Pearl-et-al-2011:high, Iess-Stevenson-Parisi-et-al-2014:gravity, Porco-Spitale-Mitchell-et-al-2007:enceladus}. Enceladus is tidally locked to Saturn with almost zero obliquity, and so tidal heating patterns will be almost perfectly symmetric between the two hemispheres in the absence of an asymmetry in the ice shell \citep{Chen-Nimmo-2011:obliquity,Baland-Yseboodt-Van-2016:obliquity}.

  It has been suggested that hemispheric asymmetry could stem from an a-priori anomaly due to a giant impact and/or a monopole structure in geological activity \citep{Han-Showman-2010:coupled, Behounkova-Tobieb-Chobletb-et-al-2012:tidally, Rozel-Besserer-Golabek-et-al-2014:self, Roberts-Stickle-2021:breaking}. It is supposed that on geological timescales, this anomaly then moves over the south pole by true polar wandering \citep{Nimmo-Pappalardo-2006:diapir, Stegman-Freeman-May-2009:origin, Tajeddine-Soderlund-Thomas-et-al-2017:true}. However, given that currently there is only one such hot spot on Enceladus, implicit assumptions have to be made about the frequency of giant impacts and their recovery time.

\citet{Hemingway-Rudolph-Manga-2019:cascading} propose that the ice shell could be torn apart by the overpressure induced by a freezing of the subsurface ocean. Since the ice shell is thinnest and hence weakest over the poles, due to the polar-amplified tidal dissipation in the ice shell and core \citep{Beuthe-2013:spatial, Choblet-Tobie-Sotin-et-al-2017:powering}, the initial fracture would most likely occur at one of the poles. Once this happens, the pressure would be released thus preventing a similar event from happening over the other pole. The enhanced dissipation from polar venting will further thin the ice there.
  
\citet{Kang-Flierl-2020:spontaneous}, instead, suggest that any small random asymmetry in ice shell thickness, be it induced by giant impacts \citep{Roberts-Stickle-2021:breaking}, obliquity tides \citep{Tyler-2009:ocean}) or other mechanisms, can amplify on million-year timescales. The fastest growth occurs for thickness variations which have the gravest inter-hemispheric scale, significantly thinning the ice shell at one of the poles and thereby promoting fracture formation there. The hemispherically-asymmetric mode will grow and dominate over time for two reasons. First, rheology feedback makes thin regions of the ice shell weaker and more mobile which in turn generates more heat and yet more thinning \citep{Beuthe-2019:enceladuss} --- this enhances all thickness variations uniformly. Second, ice flow is more efficient at smoothing out thickness variations which have small spatial scales than those with large scales \citep{Ashkenazy-Sayag-Tziperman-2018:dynamics}. The combined action of these two processes can account, it is argued, for both the concentration of geysers over one pole and the significant hemispheric asymmetry in ice shell thickness.  

  Whatever the processes that influence the evolution of the ice shell, its freezing and melting will induce salinity changes just under the ice, and its thickness variations will strongly influence the temperature at the water-ice interface through the suppression of the freezing point of water with pressure --- see Fig.\ref{fig:EOS-Hice-Heatflux}a. Thus one can expect lateral salinity and temperature gradients at the water-ice interface, both of which will drive ocean circulation. This circulation will in turn redistribute heat \citep{Tyler-2014:comparative}, reshaping the ice shell by inducing freezing and melting \citep{Kang-Mittal-Bire-et-al-2021:how}. This interaction between the ice shell and the ocean just below it could provide an additional feedback, promoting, or otherwise, hemispheric symmetry breaking. The focus of the present work is to study the possible role of cross-equatorial ocean heat transport in enhancing or diminishing inter-hemispheric asymmetries in ice thickness. 

  \section{Explorations of cross-equatorial heat transport in a model of Enceladus' ocean circulation }
    
  We simulate the ocean circulation and its interaction with a poleward-thinning ice shell on an Enceladus-like icy satellite, introduce small inter-hemispheric asymmetries in ice thickness and explore the efficacy of cross-equatorial heat transport. The Massachusetts Institute of Technology ocean model \citep[MITgcm, ][]{MITgcm-group-2010:mitgcm, Marshall-Adcroft-Hill-et-al-1997:finite} is employed in a zonally-averaged setup similar to, but slightly different from, that described in \citep{Kang-Mittal-Bire-et-al-2021:how}. Instead of trying to understand the possible ocean circulation on Enceladus driven by the observed ice thickness variations, here we use the method of perturbations to study the feedbacks at play when the ice shell is assumed to be slightly asymmetric about the equator. Observations suggest a rather marked hemispheric asymmetry in ice thickness \citep{Iess-Stevenson-Parisi-et-al-2014:gravity, Beuthe-Rivoldini-Trinh-2016:enceladuss, Tajeddine-Soderlund-Thomas-et-al-2017:true, Cadek-Soucek-Behounkova-et-al-2019:long, Hemingway-Mittal-2019:enceladuss} but our goal here is to investigate the ocean feedbacks --- is a small asymmetry amplified or damped by ocean circulation and its associated ice-ocean interaction? Rather than the freezing/melting rate being prescribed, as in \citep{Kang-Mittal-Bire-et-al-2021:how}, here it depends on the circulation, allowing us to study feedbacks. Further details of the model configuration can be found in the supplementary materials (SM). Typically solutions take many thousands of years to reach equilibrium.
  
  \begin{figure}[htbp!]
    \centering \includegraphics[page=1,width=0.75\textwidth]{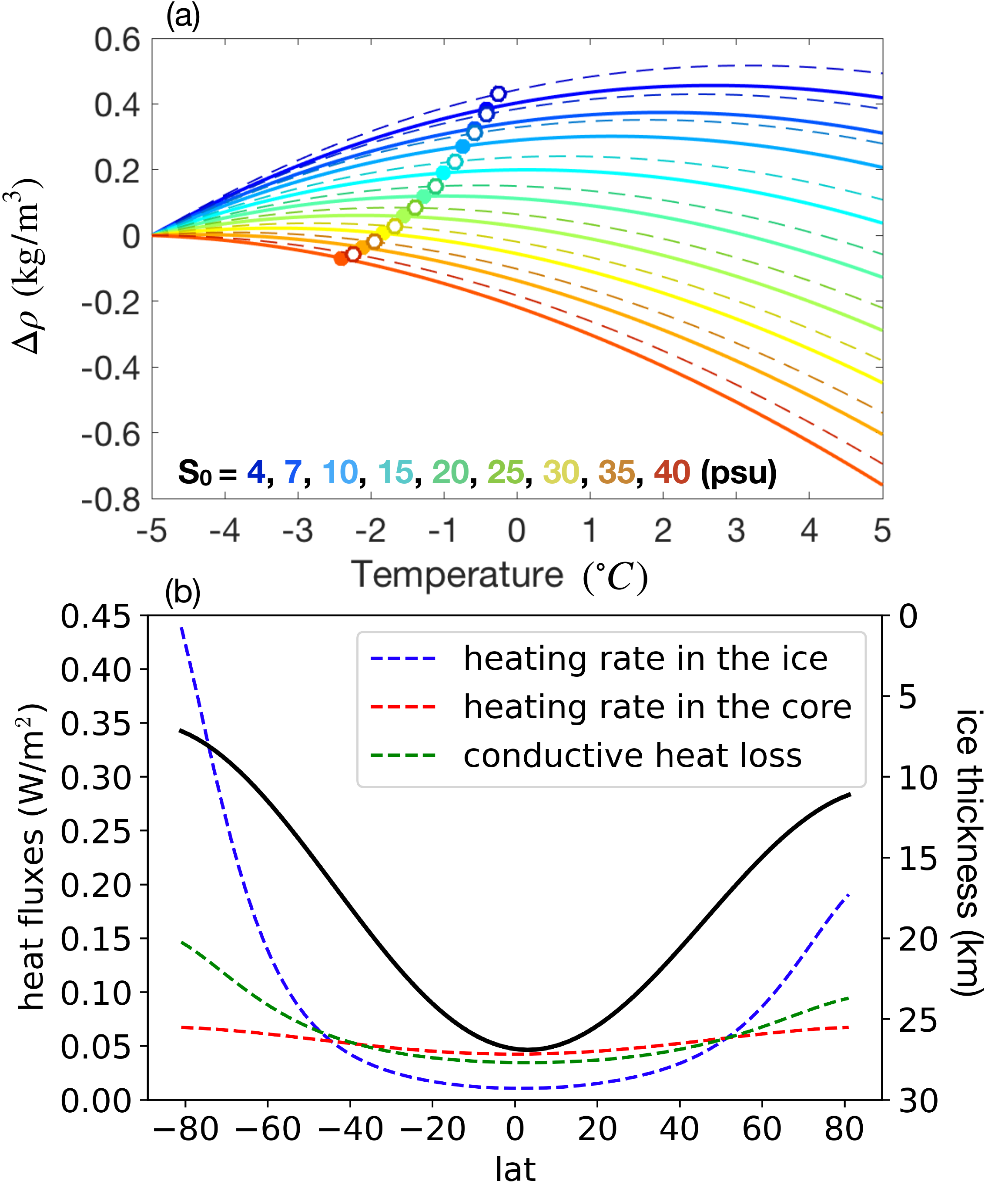}
    \caption{\small{(a) The density anomaly of water near the freezing point as a function of temperature for various salinities and two different pressures. Cold to warm colors denote increasing salinity (see legend). The pressure beneath the 26-km thick ice shell at the equator is used to compute the solid curves and the pressure beneath the 7-km south polar ice shell is used for the dashed curves. In each case the density at $-5$~$^\circ$C is chosen as a reference. The circles indicate the freezing point, full circles at the base of the equatorial ice shell and open circles at the base of the south polar ice shell: note the rightward shift of the open circles, which indicate the freezing point increases moving from equator to pole. (b) The ice shell thickness assumed here (solid curve, axis on the rhs) plotted as a function of latitude, along with the corresponding profiles of tidal heating in the ice shell (blue dashed, axis on the lhs), in the core (red dashed) and heat loss by conduction through the ice shell (green dashed). The global-mean of tidal heating in the core and the ice shell are set equal to the global-mean conductive heat loss.}}
    \label{fig:EOS-Hice-Heatflux}
  \end{figure}
  
  The assumed ice thickness profile, which thins toward the poles and is slightly asymmetric about the equator --- thinner over the south pole than the north --- is shown by the black solid line in Fig.\ref{fig:EOS-Hice-Heatflux}b. The suppression of the freezing point on increasing pressure leads to temperature variations beneath the ice which induces a circulation. This redistributes heat, and exchanges with the ice modulate the freezing/melting (as set out in detail in Eqs.~13, 14, SM). This in turn changes the salinity through brine rejection/freshwater production (Eq.~16, SM). In our model, however, the ice shell geometry is not allowed to evolve under the assumption that lateral ice flow maintains balance \footnote{We expect the results to remain similar with ice shell evolution on, and during the period of integration, we don't expect significant change of ice geometry due to the long adjusting timescale.}. Meanwhile, the core may also be a source of heat due to its tidal flexing, and is transmitted to the ocean through a bottom heat flux. Such core heating has been argued to be necessary to keep Enceladus from freezing up, given that the ice dissipation is likely insufficient at the present-day eccentricity \citep{Beuthe-2019:enceladuss, Choblet-Tobie-Sotin-et-al-2017:powering}. According to Sandstrom's theorem \citep[][Malte Jansen, private communication]{Wunsch-2005:thermohaline}, both temperature and salinity forcings from the ice and the core can drive ocean circulation if buoyancy sinks occur higher up the water column than buoyancy sources. Furthermore, vertical diffusion may provide additional energy to the system by pumping dense water upward \citep{Young-2010:dynamic}.
  
The normalized heat fluxes due to tidal dissipation in the shell and the core, together with the conductive heat loss through the ice shell, are shown in Fig.~\ref{fig:EOS-Hice-Heatflux}b. There the global-mean of the heat fluxes in the core and the ice shell are set equal to the global-mean conductive heat loss. In our experiments, the core/ice shell heat fluxes are also adjusted to guarantee that they exactly balance the globally-averaged conductive heat loss. Since regions with a thinner ice shell tend to be more dissipative (due to the ice rheology feedback) \citep{Beuthe-2019:enceladuss}, heat production in the ice shell is concentrated over poles, and especially the south pole. Heating in the core, by contrast, is not affected by the ice thickness and is therefore assumed to be symmetric between the two hemispheres, as shown in Fig.~\ref{fig:EOS-Hice-Heatflux}b. 
  
  Under hemispherically-asymmetric forcing, a cross-equatorial overturning circulation forms in the ocean. If heat is carried from the northern hemisphere to the southern hemisphere by ocean circulation, the hemispheric asymmetry of ice thickness is enhanced, and vice versa. Given the limited knowledge of Enceladus' ocean salinity \citep{Postberg-Kempf-Schmidt-et-al-2009:sodium, Zolotov-2007:oceanic, zolotov2014can, Glein-Postberg-Vance-2018:geochemistry, Ingersoll-Nakajima-2016:controlled, Hsu-Postberg-Sekine-et-al-2015:ongoing, Kang-Mittal-Bire-et-al-2021:how} and the partition of heat production between the ice shell and the silicate core \citep{Choblet-Tobie-Sotin-et-al-2017:powering, Beuthe-2019:enceladuss, Kang-Bire-Campin-et-al-2020:differing}, a series of experiments were carried out covering a range of ocean salinities (5, 10, 15, 20 and 25~psu) and core/shell heat partitions (0\%/100\%, 20\%/80\%, 40\%/60\%, 60\%/40\%, 80\%/20\%, 100\%/0\%). Hereafter, we use S$x$s$y$ to refer to the experiment with a $x$~psu salinity and $y\%$ of heat generated in the shell, $s$. It should be noted that the ice shell heating rates used in scenarios in which it is the dominant heat source is perhaps 1 order of magnitude higher than the predictions of tidal dissipation models, given our current understanding of ice rheology \citep{Beuthe-2019:enceladuss}. Although more advanced models of ice rheology can lead to more heat production \citep{Gevorgyan-Boue-Ragazzo-et-al-2020:andrade}, thus far, the attempt to fully account for purported heat losses has not yet been successful --- see \cite{Robuchon-Choblet-Tobie-et-al-2010:coupling, Shoji-Hussmann-Kurita-et-al-2013:ice, Behounkova-Tobie-Choblet-et-al-2013:impact, McCarthy-Cooper-2016:tidal, Beuthe-2019:enceladuss, Soucek-Behounkova-Cadek-et-al-2019:tidal, Gevorgyan-Boue-Ragazzo-et-al-2020:andrade}.


  \begin{figure}[htbp!]
    \centering \includegraphics[page=12,width=0.95\textwidth]{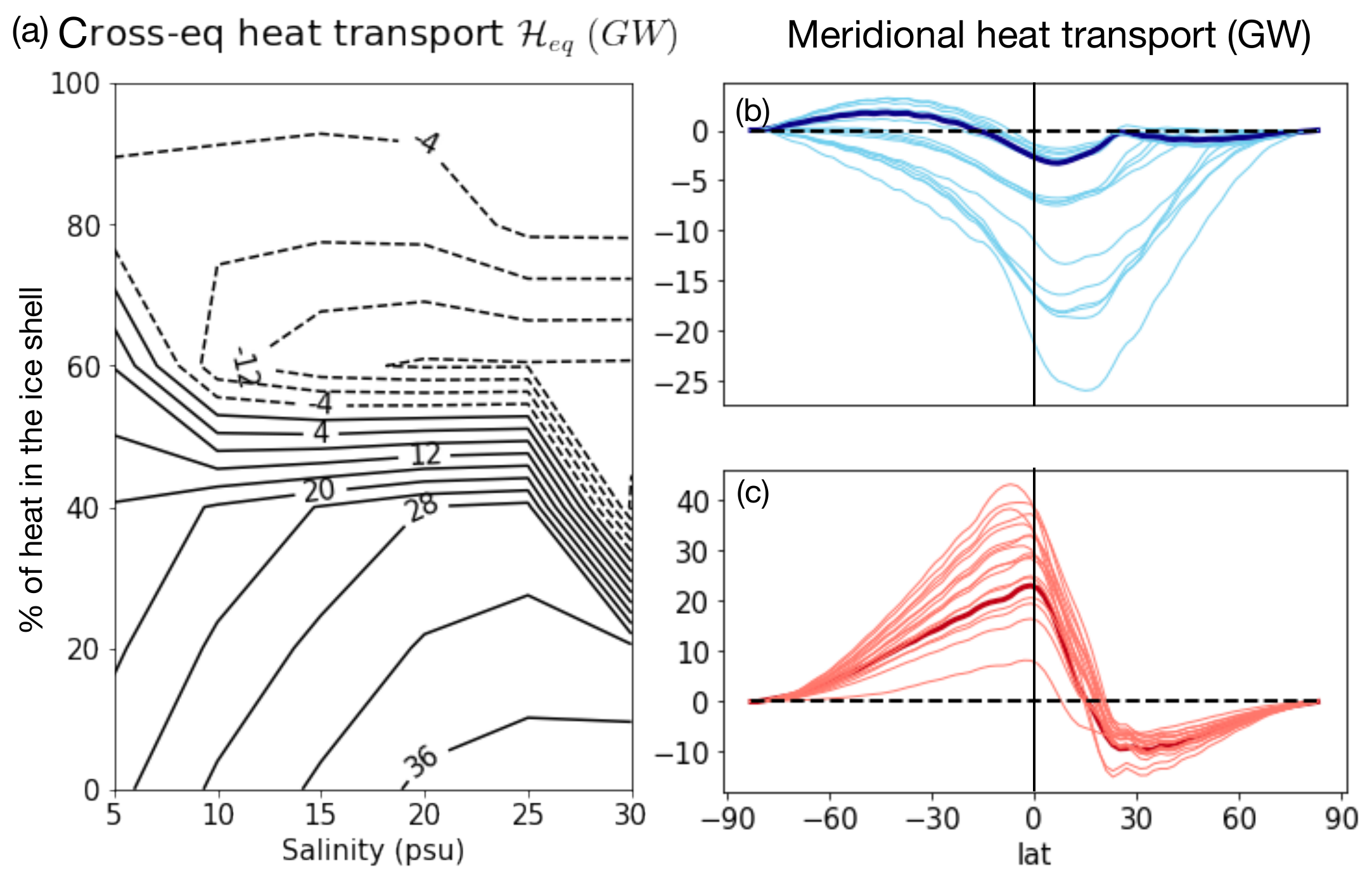}
    \caption{\small{(a) Heat transport across the equator, $\mathcal{H}_{\mathrm{eq}}$, from a range of experiments assuming various heat partitions and ocean salinities: dashed lines are negative and continuous lines are positive. $\mathcal{H}_{\mathrm{eq}}$ is southward/northward when heating dominates in the ice shell/core. Panel (b,c) show meridional heat transport profiles for experiments with southward (in red) and northward (in blue) cross-equatorial heat transport, respectively. Among all profiles, we use a thicker and darker curve to show the case with the highest salinity and shell heating (S30s80), and with the lowest salinity and shell heating (S5s0). These two solutions, representative of northward and southward cross-equatorial heat transport, are shown in more detail in Fig.~\ref{fig:climatology-2-cases}.}}
    \label{fig:meridional-heat-transport}
  \end{figure}

A key finding of our study is summarized in Fig.\ref{fig:meridional-heat-transport}(a), where cross-equatorial heat transport is shown for various ocean salinities and core-shell heat partitions. We see that when most of the heating is assumed to be in the ice shell, ocean circulation tends to carry heat from the northern hemisphere to the southern hemisphere. This induces melting in the southern hemisphere where the ice shell is already thin, enhancing hemispheric asymmetry. The opposite happens when heating is assumed to dominate in the core. Ocean salinity plays a secondary role through its affect on the sign and magnitude of the thermal expansion coefficient, the slope of the curves shown in Fig.\ref{fig:EOS-Hice-Heatflux}a. 
  

  We use the direction of equatorial heat transport as a criterion to separate our experiments into two groups, one which enhances hemispheric asymmetry in the ice shell (negative/southward cross-equatorial heat transport) and the other that suppresses it (positive/northward cross-equatorial heat transport). Panel (b) and (c) in Fig.\ref{fig:meridional-heat-transport} show the meridional heat transport profiles for the these two groups, respectively. Conspicuously, the pattern of heat transport profiles are broadly the same within each group, attesting to the similarity of their ocean circulation and thermodynamic states (not shown). The solution with the highest salinity and shell heating (S30s80 ---  asymmetry-enhancing), and the lowest salinity and core heating (S5s0 --- asymmetry-suppressing) are shown in more detail in Fig.\ref{fig:climatology-2-cases}.

  Both groups converge heat toward the equator, but the amplitude is much stronger in the asymmetry-suppressing group which have the preponderance of heating coming from the core. Since the global heat flux lost by the moon is around 35-40~GW \citep{Beuthe-2019:enceladuss}, convergence of such large quantities of heat toward the equator, as is typical of asymmetry-suppressing scenarios, would ultimately lead to melting of the equatorial ice shell, rendering it impossible to sustain the observed equator-to-pole ice thickness gradient. A typical asymmetry-suppressing scenario has over 50\% of the global heat flux (20-30~GW) transported to the northern hemisphere, where the ice is thicker, sufficient to suppress the development of asymmetry. Furthermore, the observed asymmetry of Enceladus' ice shell is much stronger than prescribed here \citep{Iess-Stevenson-Parisi-et-al-2014:gravity, Beuthe-Rivoldini-Trinh-2016:enceladuss, Tajeddine-Soderlund-Thomas-et-al-2017:true, Cadek-Soucek-Behounkova-et-al-2019:long, Hemingway-Mittal-2019:enceladuss}, making the situation even more problematic for retention of the observed ice shell geometry.

  In the remainder of our paper, we describe the contrasting circulations of the asymmetry-enhancing and asymmetry-suppressing solutions and then go on to discuss the mechanisms that facilitate cross-equatorial heat transport and hence symmetry-breaking.

  \section{Contrasting ocean circulations driven by shell-dominated vs core-dominated heating}
\begin{figure}[htbp!]
    \centering \includegraphics[page=13,width=0.75\textwidth]{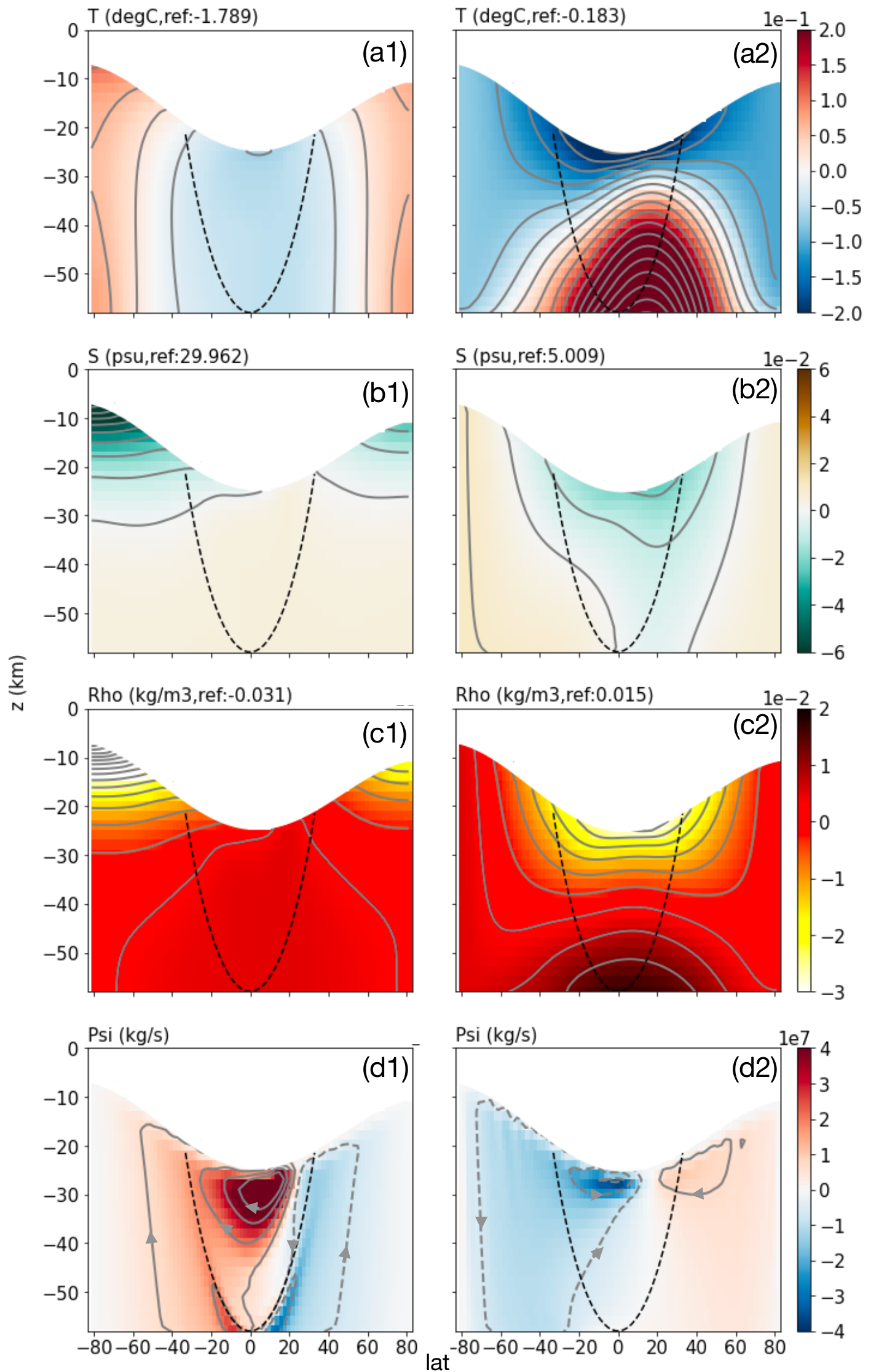}
    \caption{\small{Typical circulation and thermodynamic ocean state of (column 1) the core-heating, asymmetry-enhancing group and (column 2) the shell-heating, asymmetry-suppressing group. The asymmetry-enhancing solution has the highest salinity and shell heating (S30s80), and the asymmetry-suppressing solution has the lowest salinity and core heating (S5s0). Moving from top to bottom we present temperature $T$, salinity $S$, density anomaly $\Delta\rho$ and the meridional overturning streamfunction $\Psi$ with arrows to indicate the sense of circulation. The reference temperature and salinity (marked at the top of each plot) are subtracted from $T$ and $S$ enabling us to better capture the spatial patterns. Positive $\Psi$ indicates a clockwise overturning circulation. The contour intervals are 0.04K, 0.01~psu, 0.005~kg/m$^3$ and 1.2$\times$10$^{7}$~kg/s from top to bottom. The tangent cyclinder is shown by the dashed line in each panel.}}
    \label{fig:climatology-2-cases}
  \end{figure}

  In the ice shell-heating solutions, the highly polar-amplified heating profile (blue dashed curve in Fig.\ref{fig:EOS-Hice-Heatflux}b) induces melting over the poles (especially the south pole) and freezing elsewhere. This reduces the salinity over the polar regions compared to the equator (Fig.\ref{fig:climatology-2-cases}-b1). Temperature is higher under the thin ice shell over the poles because the freezing point is higher when the ice shell is thinner. This is manifested by the rightward shift of the open circles (which indicate the freezing point at the base of the south polar ice shell) relative to the filled circles (marking the freezing point at the base of the equatorial ice shell) in Fig.\ref{fig:EOS-Hice-Heatflux}a. When salinity is higher than 22~psu (the case shown has a salinity of 30~psu), the thermal expansion coefficient of water (given by the slopes of the curves in Fig.\ref{fig:EOS-Hice-Heatflux}a) is positive. Thus both high temperature and low salinity make polar waters lighter than equatorial waters (Fig.\ref{fig:climatology-2-cases}-c1). This induces downwelling in low latitudes and upwelling in high latitudes (Fig.\ref{fig:climatology-2-cases}-d1). Through mixing of warm polar water with cold equatorial water, heat is converged toward the equator (Fig.\ref{fig:meridional-heat-transport}b).
  
  
  If most heat is generated in the core, the heating profile is oblivious to the poleward-thinning ice shell geometry but, nevertheless, is slightly polar-amplified (red dashed curve in Fig.\ref{fig:EOS-Hice-Heatflux}b). Instead of being directly transmitted to the ice shell, this moderate polar amplification in heating is overwhelmed by the equatorward heat convergence (Fig.\ref{fig:meridional-heat-transport}c), leading to melting of ice at low latitudes and freezing at high latitudes. As a result, the salinity of equatorial (polar) waters is brought down (up), as can be seen in Fig.\ref{fig:climatology-2-cases}-b2. Since the case shown has a salinity of 5~psu, lower than the 22~psu transition point, water expands upon cooling (Fig.\ref{fig:EOS-Hice-Heatflux}a). Thus, in contrast to the shell-heating group, both high temperature (Fig.\ref{fig:climatology-2-cases}-a2) and high salinity (Fig.\ref{fig:climatology-2-cases}-b2) make polar waters denser rather than lighter than equatorial waters. This drives an ocean circulation in the opposite direction to the shell-heating case, with sinking in high latitudes and upwelling in low latitudes (Fig.\ref{fig:climatology-2-cases}-d2). The equatorward flow in the deeper ocean picks up heating from the seafloor, should it be present, carrying heat toward the equator. This is in addition to equatorward heat transport associated with mixing between polar water and equatorial water. The net result is a much stronger convergence of heat to the equator in this asymmetry-suppressing group compared to the asymmetry-enhancing group (Fig.\ref{fig:meridional-heat-transport}c).
  
It should be noted that heat is converged toward low latitudes in all experiments, just as described in \citet{Kang-Mittal-Bire-et-al-2021:how}. This indicates that the ice thickness gradient between equator and poles will always be reduced by ocean circulation (Fig.\ref{fig:meridional-heat-transport}b,c). If this cannot be counterbalanced by polar-amplified tidal heating in the ice, the equator-to-pole thickness gradient will erode over time, leading to a state which would be inconsistent with the observed ice geometry \citep{Iess-Stevenson-Parisi-et-al-2014:gravity, Beuthe-Rivoldini-Trinh-2016:enceladuss, Tajeddine-Soderlund-Thomas-et-al-2017:true, Cadek-Soucek-Behounkova-et-al-2019:long, Hemingway-Mittal-2019:enceladuss}.

Finally, we note that not only are the overturning circulations of the two solutions of opposite sign, but they are also slightly shifted relative to the equator, one north the other south. It is this shift that leads to cross-equatorial transport of heat and the possibility of symmetry-breaking in ice thickness, as we now discuss.

   \begin{figure}[htbp!]
    \centering \includegraphics[page=14,width=0.7\textwidth]{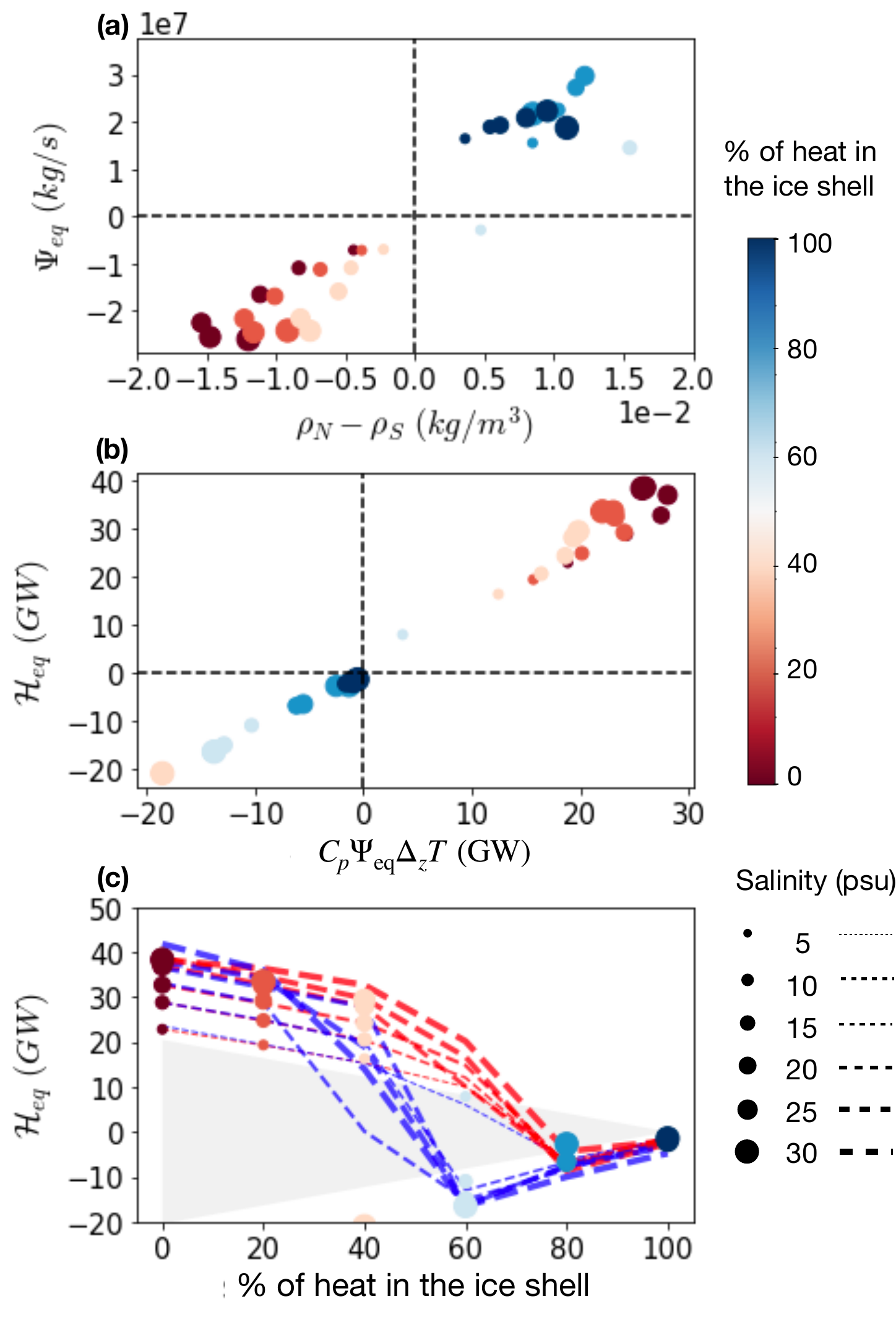}
    \caption{\small{(a) The correlation between the cross-equator overturning circulation $\Psi_{\mathrm{eq}}$ (the vertically averaged overturning streamfunction at the equator) and the density difference, $\rho_N-\rho_S$, between the north pole and south pole. (b) The proportional relationship between the meridional heat transport across the equator $\mathcal{H}_{eq}$ and the product of $\Psi_{\mathrm{eq}}\Delta_z T$, where $\Psi_{\mathrm{eq}}$ is the cross-equatorial mass exchange, and $\Delta_z T$ is the top-minus-bottom temperature difference at the equator, which is typically negative. Dots in (c) show the cross-equatorial heat transport $\mathcal{H}_{eq}$ as a function of core/shell heat partition. In all panels, each dot represents one experiment with a specific ice-core heat partition (denoted by the color of the dot) and ocean salinity (denoted by the size of the dot). The grey shading in panel (c) delineates the possible ranges of cross-equatorial heat transport if only the bottom heating was redistributed. Dashed curves shown in panel (c) are the equilibrium $\mathcal{H}_{eq}$ for two groups of experiments initialized from different initial conditions. Red curves show the equilibrium $\mathcal{H}_{\mathrm{eq}}$ against the core/shell heat partition for experiments initialized by the equilibrium fluid state of the experiment with 100\% core-heating. Blue curves show the same but for experiments initialized using experiment with 80\% heat generated in the ice shell. }}
    \label{fig:Hy-Psi-dRho-correlation}
  \end{figure}

  \section{Mechanisms behind cross-equatorial heat transport and symmetry breaking}
Ocean circulation can enhance or suppress symmetry breaking by transporting heat across the equator. This can be facilitated by a circulation that straddles the equator connecting the two hemispheres together, much as our earth's ocean's overturning circulation connects the northern and southern hemispheres, resulting in cross equatorial heat transport in the Atlantic sector --- see, for example, the review by \citet{Buckley-Marshall-2016:observations}. On earth, the ocean's are confined in basins defined by coasts that cross the equator and support western boundary currents. These boundary currents facilitate cross equatorial transport, for example in the giant meridional overturning circulation of the Atlantic ocean. In our simulations of Enceladus, there are no such topographic assists to cross-equatorial transport because, in the absence of observational constraints, our model assumes a flat bottom. Instead, the top and bottom boundaries provide the necessary friction and ageostrophic flow to support an overturning circulation. Due to the hemispheric asymmetry in the forcing, the overturning cells move (slightly) away from the equator, leading to cross-equatorial heat transport. In the asymmetry-enhancing cases, water crosses the equator in the interior ocean and returns near the surface (Fig.\ref{fig:climatology-2-cases}-d1). The opposite is true in the asymmetry-suppressing cases (Fig.\ref{fig:climatology-2-cases}-d2). We now build a simple model of the cross-equatorial heat transport.

The cross-equatorial heat transport, $\mathcal{H}_{\mathrm{eq}}$, is jointly controlled by the inter-hemispheric mass exchange, which can be measured by the vertically averaged mass-transport streamfunction at the equator, $\Psi_{\mathrm{eq}}$, and by the top-minus-bottom temperature difference $\Delta_z T$. As can be seen in Fig.\ref{fig:Hy-Psi-dRho-correlation}b, the diagnosed cross-equatorial heat transport correlates very nicely with $C_p\Psi_{\mathrm{eq}}\Delta_z T$ (here $C_p=4000$~J/kg/
K is the heat capacity of water). When heat production is predominantly in the core (denoted by reddish dots) equilibrium solutions are spread in the first quadrant of the diagram, which corresponds to energy gain in the northern hemisphere; when heat production is in the ice shell (bluish dots) solutions congregate in the third quadrant with southward heat transport at the equator. In all experiments, temperature increases with depth at the equator ($\Delta_z T<0$): for those with core heating, the bottom water is naturally warmer; for those without, $|\Delta_z T|$ is much smaller but remains negative because the temperature at the equatorial water-ice interface is close to the local freezing point, which is strongly depressed due to the thick ice shell there. We see then that the direction of inter-hemispheric heat transport is controlled by the sign of $\Psi_{\mathrm{eq}}$.

 The cross-equatorial overturning, $\Psi_{\mathrm{eq}}$, is driven by the density difference between the two hemispheres (Fig.\ref{fig:climatology-2-cases}-c). The two correlate very well, as can be seen in the scatter plot in Fig.\ref{fig:Hy-Psi-dRho-correlation}a of $\Psi_{\mathrm{eq}}$ vs the pole-to-pole density difference $\rho_N-\rho_S$. This strong correlation is likely to stem from the positive feedback between the two variables: if water in the northern hemisphere is generally denser than in the southern hemisphere ($\rho_N-\rho_S>0$), a positive $\Psi_{\mathrm{eq}}$ will be induced. Given that $\Delta_z T<0$, ocean circulation will then transport heat into the southern hemisphere, facilitating melting there and further increasing the density contrast $\rho_N-\rho_S$. Ocean salinity affects the strength of this positive feedback, by regulating the amount of salinity change induced per unit amount of freezing/melting. This is why high salinity experiments (denoted by dots of larger size) are found farther away from the origin in Fig.\ref{fig:Hy-Psi-dRho-correlation}a. In the presence of a positive feedback, the cross-equatorial circulation and heat transport, once activated, will amplify over time until being stabilized --- as $\Psi_{\mathrm{eq}}$ strengthens, salinity gradients induced by the ice freezing/melting will be smoothed out more readily, limiting the salinity gradient that can be sustained.
  
Constrained positive feedback is known to be the key to systems which exhibit hysteresis and bistable behavior \citep{Meadows-others-1997:places}. To seek multiple equilibrium solutions for each parameter set, we ran the same experiments across our salinity and core/shell heat partitions, but instead of initializing the model with zero salinity variations and zero flow, we used the equilibrium fluid state (temperature, salinity and flow field) of experiments with 0\% shell-heating, which exhibit strong northward heat transport across the equator  and with 80\% shell-heating, which exhibit strong southward heat transport. As shown by the dashed curves in Fig.\ref{fig:Hy-Psi-dRho-correlation}c, when initialized from the equilibrium state with 0\% shell-heating (red dashed curves), more heat tends to be transported northward, in comparison to the same experiments initialized from the equilibrium state with just 80\% shell-heating (blue dashed curves). Just as in a typical hysteresis diagram, the difference between the two sets of experiments maximizes when the heat partition is around 50-50 and vanishes when the core-heat-partition or shell-heat-partition approaches 100\%. 


  The direction of the cross-equatorial circulation, and thereby the ocean feedback onto ice shell hemispheric asymmetry, is controlled by the initial pole-to-pole density difference, $\rho_N-\rho_S$, before any cross-equatorial circulation or heat transport has formed. If water in the northern hemisphere is slightly heavier, ocean circulation would transport heat to the southern hemisphere, enhancing the preexisting asymmetry; if the opposite is true, the asymmetry would be suppressed. In principle, the density can be affected by both salinity and temperature. However, salinity dominates over temperature except for the deep ocean in cases such as S5s0 (column-2 of Fig.\ref{fig:climatology-2-cases}) --- this corresponds to a fresh ocean with a significant amount of heating from the core; because of anomalous expansion, the ocean is much warmer at the bottom whilst remaining stable.

Since salinity is always the dominant factor in setting pole-to-pole density differences thanks to the positive feedback, we must understand which of the poles freezes/melts relative to the other before cross-equatorial heat transport begins to develop. There are three mechanisms that affect this ``initial'' freezing/melting: 1) conductive heat loss, $\mathcal{H}_{\mathrm{cond}}$, will be more efficient over the south pole due to the relatively thin ice there; 2) since the ice shell is thinner over the south pole, more tidal heating, $\mathcal{H}_{\mathrm{ice}}$, is produced there as a result of the ice rheology feedback \citep{Beuthe-2019:enceladuss, Kang-Flierl-2020:spontaneous}, leading to melting; 3) the ice thickness gradient is stronger in the southern hemisphere, and so will induce more equator-to-pole mixing, more equatorward heat transport due to the ice-pump mechanism \citep{Lewis-Perkin-1986:ice, Kang-Mittal-Bire-et-al-2021:how}, and hence more freezing at the south pole. The first and the last processes suppress hemispheric symmetry-breaking, whereas the second mechanism enhances symmetry-breaking. Only when the ice shell produces most of the heat, does the second mechanism dominate resulting in enhanced melting over the south pole (or less freezing) than the north pole. As time progresses the cross-equatorial ocean circulation then enhances this initial tendency. In this way, we can rationalize the ocean feedback on core/shell heat partition shown in Fig.\ref{fig:meridional-heat-transport}a.

Whether melting of ice occurs more at the north pole vs the south pole depends on the sign of (see Eq.\ref{eq:boundary-heat}):
\begin{equation}
    \label{eq:Lq}
   L\Delta{q}=\Delta\mathcal{H}_{\mathrm{cond}}-\Delta\mathcal{H}_{\mathrm{ice}}-\Delta\mathcal{H}_{\mathrm{ocn}}\nonumber
\end{equation}
where $\Delta$ denotes the difference between the quantity evaluated over the north pole and the south pole and $q$ is the freezing rate. The difference between $\mathcal{H}_{\mathrm{cond}}$ and $\mathcal{H}_{\mathrm{ice}}$ over the two poles can be quantified given that the ice over the south polar regions is about 30\% thinner than that over the north pole ($\Delta H/H=0.3$, see black solid curve in Fig.~\ref{fig:EOS-Hice-Heatflux}b). Since the conductive heat loss $\mathcal{H}_{\mathrm{cond}}$ (Eq.\ref{eq:H-cond}) is inversely proportional to the ice thickness $H$ and the ice tidal dissipation $\mathcal{H}_{\mathrm{ice}}$ (Eq.\ref{eq:H-tide}) is proportional to $H$ raised to the power $p_\alpha$ (set to $-2$ in this study), the relative melt rate is:
  \begin{eqnarray}
    \label{eq:dHcond}
    L\Delta{q}&=&-\mathcal{H}_{\mathrm{cond}}\frac{\Delta H}{H}-p_\alpha\mathcal{H}_{\mathrm{ice}}\frac{\Delta H}{H}-\Delta\mathcal{H}_{\mathrm{ocn}}\nonumber\\
    &=&-\left[\mathcal{H}_{\mathrm{cond}}+s p_\alpha\mathcal{H}_{\mathrm{ice}}^{(s=100\%)}\right]\frac{\Delta H}{H} -\Delta\mathcal{H}_{\mathrm{ocn}},
  \end{eqnarray}
where $s$ is the percentage of heat produced by the ice, the superscript $(s=100\%)$ denotes the 100\% shell-heating scenario and $\mathcal{H}_{\mathrm{ocn}}$ denotes the heat flux transmitted from the ocean to the ice. As can be read off from Fig.~\ref{fig:EOS-Hice-Heatflux}b, $\mathcal{H}_{\mathrm{ice}}^{s=100\%}$ is roughly 1.5 -- 2 times as large as $\mathcal{H}_{\mathrm{cond}}$ in high latitudes. Therefore, more freezing occurs in the north relative to the south ($\Delta{q}>0$) only if $s>30\%$ with $p_\alpha=-2$, if we neglect $\Delta\mathcal{H}_{\mathrm{ocn}}$. Since the ice pump mechanism will likely be stronger given the steeper ice topography in the south, $\Delta\mathcal{H}_{\mathrm{ocn}}$ should make a positive contribution. This implies that the transition from asymmetry-suppressing to asymmetry-enhancing might occur at even higher $s$, aligned with our results (Fig.\ref{fig:meridional-heat-transport}a).

  Typically, once cross-equatorial heat transport has commenced, it strengthens over time whilst remaining the same sign, until an equilibrium is achieved. The absolute value of $\mathcal{H}_{\mathrm{eq}}$ increases as more heat is generated in the core (Fig.\ref{fig:Hy-Psi-dRho-correlation}c). This is because $\Delta_z T$ becomes more negative because the ocean bottom is heated more in the presence of a stronger bottom heat flux. A consequence is that the strongest enhancement of asymmetry does not occur when heating is all produced in the shell. Instead, it occurs when the core makes a minor but non-negligible contribution to the total heat budget. As a by-product of the aforementioned positive feedback, in most cases the amplitude of $\mathcal{H}_{\mathrm{eq}}$ exceeds the total hemispheric core heat production (grey shading in Fig.\ref{fig:Hy-Psi-dRho-correlation}c), especially when there is significant core heat production. Therefore, although bottom heating makes $\Delta_z T$ more negative and affects $|\mathcal{H}_{\mathrm{eq}}|$, it is the ocean-ice heat exchange that predominantly supports and balances the meridional heat transport. 


    \begin{figure}[htbp!]
    \centering \includegraphics[page=15,width = \textwidth]{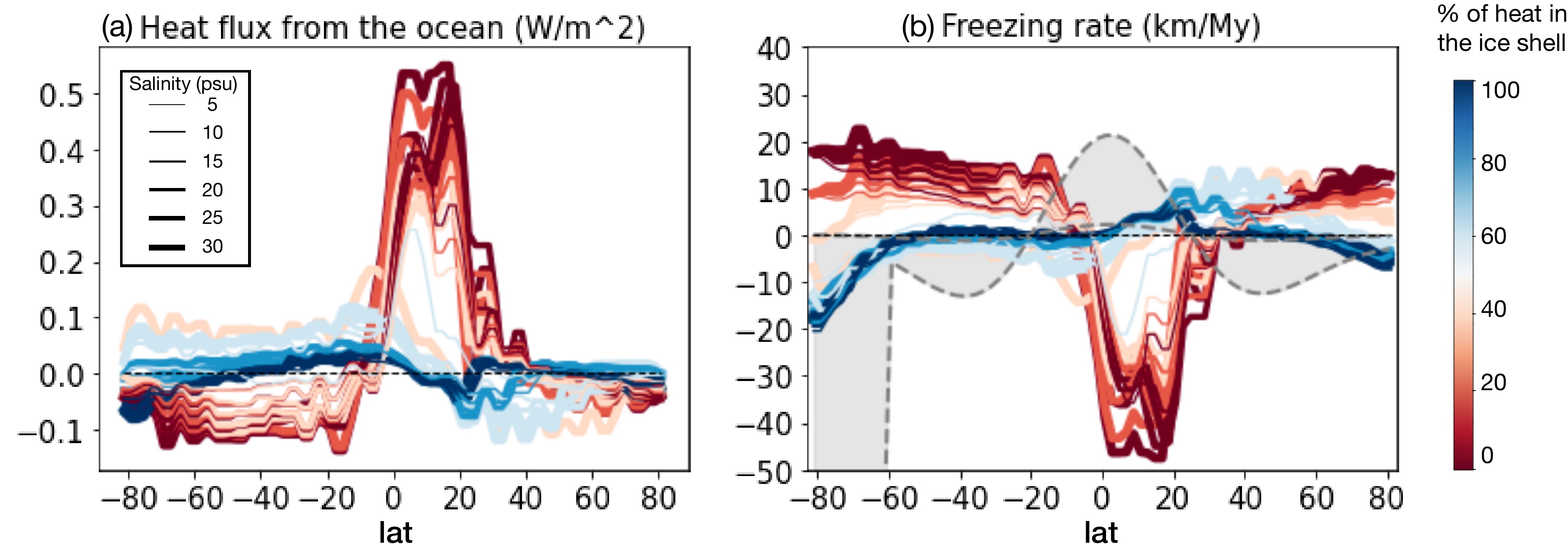}
    \caption{\small{ The ocean-ice heat exchange (left) and freezing rates (right) as a function of salinity (denoted by the line thickness) and core-shell heat partitions (denoted by the line color). The freezing/melting pattern required to balance the ice flow given the prescribed ice geometry is shown by two gray dashed curves, assuming two different ice viscosities ($\eta_{\mathrm{melt}}=10^{13},\ 10^{14}$~Pa$\cdot$s). Details of the ice flow model is given in appendix B. Since geysers over the south pole can remove significant amount of energy, and this process is not accounted for here, we set the ``lower bound'' of the freezing rate (corresponding to $\eta_{\mathrm{melt}}=10^{13}$~Pa$\cdot$s) to negative infinity to the south of 60S. The gray shaded area in between is a measure of uncertainty.}}
    \label{fig:freezing-rate}
  \end{figure}
 
Knowing the ocean-ice heat exchange (Fig.~\ref{fig:freezing-rate}a), tidal dissipation in the ice and the conductive heat loss, we can estimate the freezing/melting pattern for each scenario (Fig.~\ref{fig:freezing-rate}b). This allows us to infer ice shell evolution by comparing this freezing rate with the thickness tendencies associated with the ice flow (the likely range is shown by the gray shading in Fig.~\ref{fig:freezing-rate}), which transports ice from thick ice regions to thin ice regions under the influence of pressure gradient force (details of the ice flow model are given in appendix B). If the freezing rate is above the shaded zone, ice will grow and vice versa. In the shell-heating scenarios (shown by blueish curves), the hemispheric asymmetry will be enhanced. That said, the ocean heat transport mostly affects the low latitudes and mid-latitudes, inducing freezing around 20-60N and melting around 20S-20N. However, over geological timescales, the resultant thickness anomalies may be spread to the poles by ice flow. In contrast, the core-heating scenarios induce asymmetry which declines over time, due in part to the already weak ice-rheology feedback but also due to strong northward ocean heat transport. Moreover, because of the strong equatorward heat convergence in the core-heating scenarios \footnote{Equatorward heat convergence melts the equatorial ice shell, reducing the local salinity and hence density, which in turn facilitates upward motions in the low latitudes. This will further enhance the equatorward heat convergence by bringing the bottom hot water toward equatorial ice shell. This cannot happen in experiments with prescribed freezing/melting rate as in \citet{Kang-Mittal-Bire-et-al-2021:how}.}, the equatorial (polar) ice shell will inevitably thin (thicken) over time, aided by poleward ice flow directed down the thickness gradient. This, together with the asymmetry suppression observed in the core-heating scenarios, suggests to us that the source of heat on Enceladus comes predominantly from its ice shell rather than core. Hydrothermal activity at the sea floor may, of course, be an important local energy source for circulation (and life) but, just as on earth, it may not account for a large portion of the total heat source on Enceladus.



  \section{Summary of mechanisms and conclusions}
We have found that heat transport by ocean circulation enhances hemispheric symmetry-breaking in ice thickness when heating primarily occurs in the ice shell driven by ocean tides, whilst if heating is primarily from the core, the ocean circulation acts as a damping of ice thickness asymmetry. 

\begin{figure}[htbp!]
    \centering \includegraphics[page=8,width =0.9 \textwidth]{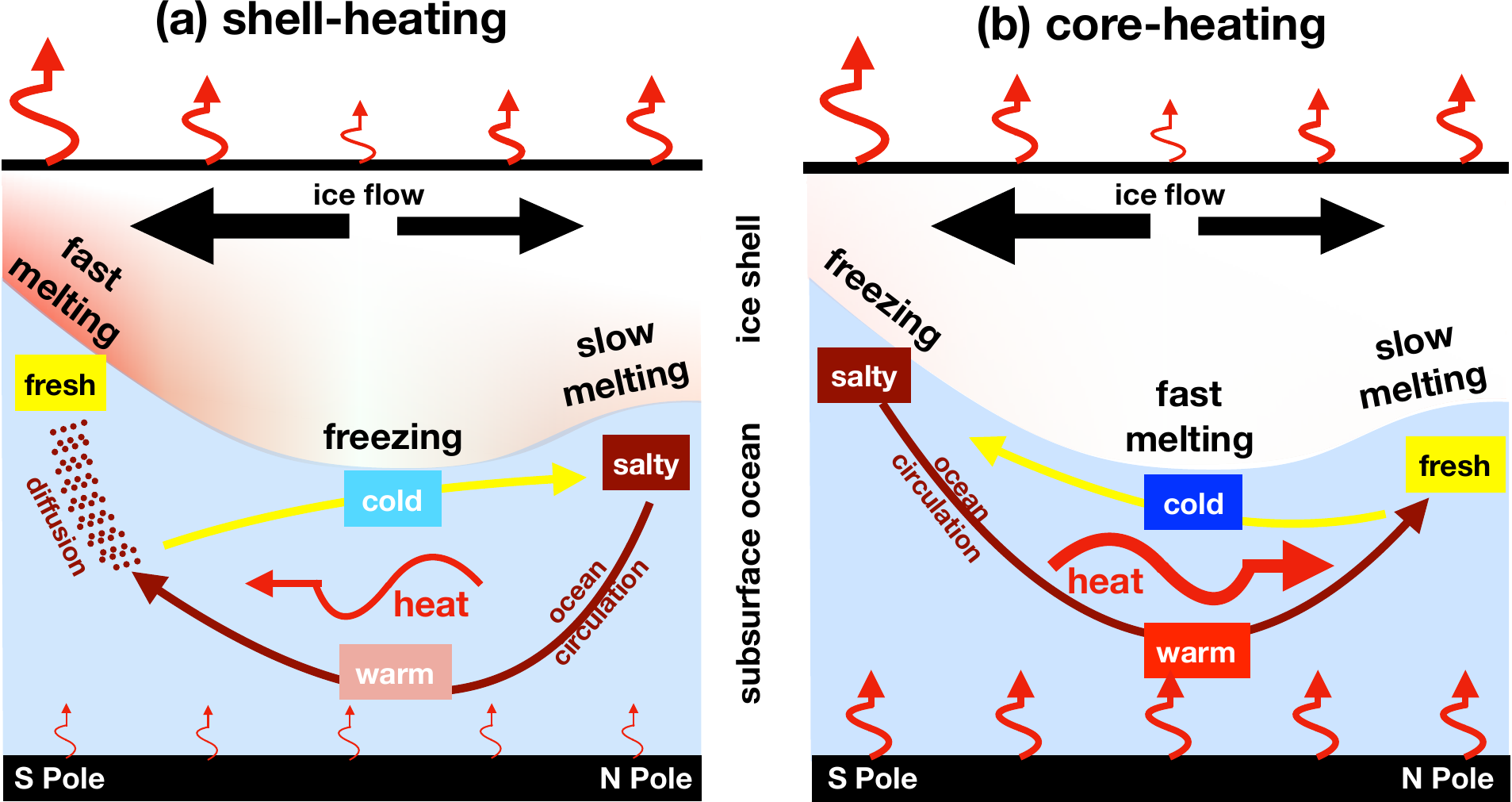}
    \caption{\small{Schematics of ocean circulation and heat transport for the (a) shell-heating and (b) core-heating scenarios. Note that only the hemispherically-asymmetric patterns are shown. The upward-pointing curly red arrows show fluxes entering from the bottom and exiting at the top. The horizontally-directed curly red arrow shows the lateral heat transport by ocean circulation. The red shading in the ice shell mark the patterns of tidal heating which peak over the poles where the ice is thin, largest over the south pole. Broad patterns of melting and freezing are marked, together with the rate, along with the associated salinity anomalies. Melting and freezing are balanced by lateral ice transport within the ice shell, directed from thick to thin. Rates of melting (fast, slow) are based on consideration of Fig.~\ref{fig:freezing-rate}b. The broad patterns of anomalous overturning circulation in the ocean are indicated by the thick curved arrows, which are color-coded to indicate the salt transport. Dark brown denotes the flow of dense salty water, light yellow that of buoyant fresh water. The water is coldest just under the thickest ice, which occurs at the equator. The water beneath is warmer, in part due to bottom heating. Water properties in different parts of the ocean are indicated by text boxes. See the main text for details. 
      }}
    \label{fig:schematics}
  \end{figure}

The mechanisms leading to cross-equatorial heat transport are sketched in Fig.~\ref{fig:schematics}. In the shell-heating scenario, heating concentrates at the south pole (shown by red shading) because the ice shell is thinner and hence more mobile there \citep{Beuthe-2019:enceladuss, Kang-Flierl-2020:spontaneous}. The conductive heat loss is also more efficient over the south pole (the red curly arrows atop of the ice shell) but is overwhelmed by tide-induced local heating. As a result, the south pole melts relative to the north. Over time, northern hemisphere waters becomes saltier and denser than in the southern hemisphere, driving an ocean circulation that sinks in the north and rises in the south (arrows). Since waters underneath the thick equatorial ice shell are the coldest of the global ocean, due to freezing point depression, such a circulation transports heat into the southern hemisphere, further thinning the ice shell there. The ocean circulation, therefore, results in patterns of melting and freezing which enhances inter-hemispheric gradients in ice thickness. This is balanced by transport of ice directed down the thickness gradient within the ice shell.

If, instead, the silicate core were to be the dominant heat source (see Fig.~\ref{fig:schematics}b), the ocean's overturning circulation would reverse sign (arrows). This is because, even though core heat production remains symmetric between the two hemispheres, the south pole freezes more rapidly relative to the north due to the more efficient conductive heat loss there (red curly arrows at the top). With sinking now in the southern hemisphere and upwelling in the north, heat is transported northward across the equator reducing ice thickness in the north. The heat flux in the core heating case is typically much stronger than with shell-heating because the vertical temperature gradient at the equator is enhanced due to bottom heating. The ocean circulation, therefore, results in patterns of melting and freezing which reduce inter-hemispheric gradients in ice thickness. Since ice thickness gradients in the ice are also flattened by ice flow, it must be concluded that the end result is one in which ice thickness gradients have been smoothed out, at which point the ocean circulation and ice distributions would be symmetrically-disposed about the equator.

The above scenarios are not strongly affected by the ocean salinity, except that a salty ocean tends to facilitate a stronger positive feedback between the ice freezing/melting and the ocean circulation and thereby faster adjustment of ice shell geometry (see Fig.\ref{fig:freezing-rate}).

Our results suggest, then, that the observed highly asymmetric Enceladean ice shell geometry \citep{Iess-Stevenson-Parisi-et-al-2014:gravity, Tajeddine-Soderlund-Thomas-et-al-2017:true, Cadek-Soucek-Behounkova-et-al-2019:long, Hemingway-Mittal-2019:enceladuss} is consistent with the primary energy source for ocean circulation being in the ice shell rather than the core. Such high dissipation rates in the ice shell may potentially be achieved through multiple scenarios as suggested by \citet{Tyler-2009:ocean, Tyler-2011:tidal, Tyler-2014:comparative, Tyler-2020:heating}, although current tidal models have difficulty achieving the high ice dissipation rate assumed here \citep{Robuchon-Choblet-Tobie-et-al-2010:coupling, Shoji-Hussmann-Kurita-et-al-2013:ice, Behounkova-Tobie-Choblet-et-al-2013:impact, McCarthy-Cooper-2016:tidal, Beuthe-2019:enceladuss, Soucek-Behounkova-Cadek-et-al-2019:tidal, Gevorgyan-Boue-Ragazzo-et-al-2020:andrade}.

For icy moons of size greater than Enceladus, most notably Europa, we expect the ocean circulation and heat transport to increase as gravity and hence surface forcing become stronger. This is because the pressure gradient at the ocean-ice interface will increase with gravity, and so give rise to greater shifts of freezing point, stronger temperature gradients and hence drive stronger ocean circulations. As a result, the role of ocean circulation may be to flatten out any equator-to-pole ice shell gradients on icy moons of larger size. Moreover, asymmetry may also be suppressed because of more efficient equator-to-pole heat transport (i.e., greater $\mathcal{H}_{\mathrm{ocn}}$ in Eq.\ref{eq:dHcond}) and a weakening of the ice-rheology feedback (smaller $\mathcal{H}_{\mathrm{ice}}/\mathcal{H}_{\mathrm{cond}}$ in Eq.\ref{eq:dHcond}).

Finally, we should remind ourselves of the many caveats that accompany our study. First and foremost, to enable us to integrate our model long enough to reach equilibrium (many thousands of years), we use a zonally symmetric configuration in which transport by baroclinic eddies is missing. We note that meridional potential vorticity gradients change sign over the water column of our solutions (not shown), and so, by the Charney-Stern criterion \citep{Charney-Stern-1962:stability}, we might expect baroclinic instability to occur. These can facilitate jet formation and affect heat and tracer transport \citep{Schneider-2006:general}. Their impact on cross-equatorial heat transport and the general circulation of icy moons is an important topic for future study. Moreover, convection is parameterized in our model rather than resolved, and the bottom heating is assumed to be uniformly distributed, albeit slightly poleward-amplified (red dashed curve in Fig.\ref{fig:EOS-Hice-Heatflux}b). Convective plumes may significantly affect heat transport, especially if the heat flux becomes geographically localised and more intense, as suggested by \citet{Choblet-Tobie-Sotin-et-al-2017:powering}. We also leave such issues for future study.

  \section*{Acknowledgments}
  This work was carried out in the Department of Earth, Atmospheric and Planetary Science (EAPS) in MIT. WK acknowledges support as a Lorenz/Houghton Fellow supported by endowed funds in EAPS. JM and SB acknowledge part-support from NASA Astrobiology Grant 80NSSC19K1427 “Exploring Ocean Worlds”. The supporting information provides detailed description of our model setup. Code and data is available upon reasonable request.

  \appendix

\section{Description of Ocean General Circulation Model}

To explore whether ocean circulation enhances or suppresses hemispheric asymmetries in the thickness of the ice shell of Enceladus, we model the large-scale overturning circulation in the Enceladean ocean encased by a hemispherically-asymmetric ice shell, and diagnose the resulting cross-equatorial heat transport. Given our uncertain knowledge of the mean salinity of the ocean and the partitioning of heat production in the core and the shell, we explore many (in fact 30) plausible combinations. The simulations are carried out using the state-of-the-art Massachusetts Institute of Technology ocean circulation model \citep[MITgcm, ][]{MITgcm-group-2010:mitgcm, Marshall-Adcroft-Hill-et-al-1997:finite} in a two-dimensional configuration (latitude and depth). The efficiency of the 2D setup enables us to integrate out the simulations to full equilibrium and explore a wide parameter space. Unless otherwise stated, initial current and salinity variations are set to zero and the initial potential temperature at each latitude set equal to the freezing point at the water-ice interface. From this initial state, each simulation is launched for 75,000 years, by which time equilibrium is reached.

The model integrates the non-hydrostatic primitive equations governing the motion of a fluid in height coordinates including a full treatment of all components of the Coriolis force. These components are typically ignored when simulating Earth's ocean because of its small aspect ratio (the ratio between depth and horizontal scale of the ocean basin), but is crucial for Enceladus's ocean, whose aspect ratio is instead order $40$km$/252$km$ \sim0.16$ and so not small. The size of each grid cell shrinks with depth due to spherical geometry and is accounted for by switching on the ``deepAtmosphere'' option of MITgcm. Since the depth of Enceladus's ocean is comparable to its radius, the variation of gravity with depth is significant. The vertical profile of gravity in the ocean and ice shell is given by, assuming a bulk density of $\rho_{\mathrm{out}}=1000$~kg/m$^3$:
  \begin{equation}
    \label{eq:g-z}
    g(z)=\frac{G\left[M-(4\pi/3)\rho_{\mathrm{out}}(a^3-(a-z)^3)\right]}{(a-z)^2}.
  \end{equation}
  In the above equation, $G=6.67\times10^{-11}$~N/m$^2$/kg$^2$ is the gravitational constant and $M=1.08\times 10^{20}$~kg and $a=252$~km are the mass and radius of Enceladus.
  
Since it takes many tens of thousands of years for our solutions to reach equilibrium, we employ a moderate resolution of $2$~degree ($8.7$~km) and run in a 2D, zonal-average configuration whilst retaining full treatment of Coriolis terms. By so doing, zonal variations are omitted. In the vertical direction, the $60$~km ocean-ice layer is separated into $30$ layers, each of which is $2$~km deep. The ocean is encased by an ice shell with meridionally-varying thickness using MITgcm's ``shelfice'' and ice ``boundary layer'' module \citep{Losch-2008:modeling}. We employ partial cells to better represent the ice topography: water is allowed to occupy a fraction of the height of a whole cell with an increment of 10\%.

  We set the ice thickness $H$ to be slightly asymmetric between the two hemispheres and thin towards each pole:
  \begin{equation}
    \label{eq:ice-thickness}
    H(\phi)=H_0+H_2P_2(\phi)+H_1P_1(\phi),
  \end{equation}
  where $H_0=20.8$~km is the mean ice thickness. The constants $H_2=-12.1$~km and $H_1=2$~km set the equator-to-pole and pole-to-pole thickness variations, respectively and $P_1$, $P_2$ are the first and second Legendre Polynomials. The thickness variation is plotted in the thick black solid curve in Fig.~\ref{fig:EOS-Hice-Heatflux}b. 
  
  We explore a wide range of background salinities from $4$~psu to $40$~psu, and adopt a linear equation of state (EOS, which determines how density depends on temperature, salinity and pressure) to make the dynamics as transparent as possible. The dependence of potential density $\rho$ on potential temperature $\theta$ and salinity $S$ is set equal to:
  \begin{eqnarray}
    \label{eq:EOS-linear}
  \rho(\theta,S)&=&\rho_0\left[1-\alpha_T(\theta-\theta_0)+\beta_S
                    (S-S_0)\right]\\
    \rho_0&=&\rho(\theta_0,S_0).\label{eq:rho0}
  \end{eqnarray}
  Here, $\rho_0,\ \theta_0$ and $S_0$ are the reference potential density, potential temperature and salinity, and $\alpha_T$ and $\beta_S$ are the thermal expansion and haline contraction coefficients respectively, given by the first derivative of the density with respect to potential temperature and salinity at the reference point, obtained using the Gibbs Seawater Toolbox \citep{McDougall-Barker-2011:getting}. We carried out two test experiments (one with $S_0=10$~psu and the other with $S_0=20$~psu) using the full ``MDJWF'' equation of state \citep{McDougall-Jackett-Wright-et-al-2003:accurate} and obtained almost identical results.

  $S_0$ is prescribed a constant value between $5$~psu and $30$~psu. $\theta_0$ is set to be the freezing temperature at $S_0$ and $P_0=2.2\times10^6$~Pa (this is the pressure under a 20.8~km thick ice sheet on Enceladus). Generally, the freezing point $T_f$ depends on local pressure $P$ and salinity $S$ following
  \begin{equation}
    \label{eq:freezing-point}
    T_f(S,P)=c_0+b_0P+a_0S,
  \end{equation}
where the constants are ascribed the following values: $a_0=-0.0575$~K/psu, $b_0=-7.61\times10^{-4}$~K/dbar and $c_0=0.0901$~degC. The pressure $P$ can be calculated using hydrostatic balance: $P=\rho_igH$ where $\rho_i=917$~kg/m$^3$ is the density of the ice and $H$ its thickness).

\subsection{Representation of subgridscale processes}

To account for the mixing of momentum, heat and salinity by turbulence that is not resolved, we set the explicit viscosity/diffusivity to be much larger than molecular values. There is no attempt to represent baroclinic instability. The default horizontal and vertical diffusivity are set to $0.001$~m$^2$/s to represent the mixing induced by high-frequency turbulence driven by tides and libration \citep{Tyler-2020:heating}. This is 3 orders of magnitude or so greater than the molecular diffusivity, but is consistent with the dissipation rate estimated by a recent study \citep{Rekier-Trinh-Triana-et-al-2019:internal}, in which both libration and tidal forcing were accounted for. According to \citet{Rekier-Trinh-Triana-et-al-2019:internal}, the tidal dissipation in the ocean is mostly induced by librations with a global dissipation rate $E$ of around 1~MW. Following the review of \citet{Wunsch-Ferrari-2004:vertical}, the vertical diffusivity can be estimated as
  \begin{equation}
    \label{eq:kappav}
    \kappa_{v}=\frac{\Gamma \varepsilon}{\rho_0 N^{2}},
  \end{equation}
where $\Gamma\sim 0.2$ is the efficiency by which kinetic energy dissipation can be used to produce potential energy, $\varepsilon=E/V$ is the ocean dissipation rate per unit area, $V=4\pi a^2D$ is the total volume of the Enceladean ocean, $\rho_0\sim 1000$~kg/m$^3$ is the density of water, $N^2=g(\partial \ln \rho/\partial z)\sim g (\Delta \rho/\rho_0)/D$ is the Brunt-Vaisala frequency. Assuming that $\Delta \rho/\rho_0$ is just $\alpha_T \Delta T_f$, where $\Delta T_f$ is the equator-to-pole temperature contrast, setting $\alpha_T\sim 1\times10^{-5}$/K and $\Delta T_f\sim0.1 K$, and substituting the above values into Eq.\ref{eq:kappav}, we find $\kappa_v\sim 0.001$~m$^2$/s. This is the default horizontal and vertical diffusivity used in our model. The diffusivity for temperature and salinity are set to be the same, so that any double diffusive effects are suppressed.

  Due to the relatively coarse model resolution, convective processes are not resolved. For regions that are convectively unstable, we overwrite the vertical diffusivity with a much greater value, $1$~m$^2$/s, to represent the vertical mixing associated with small-scale convective motions. Such approaches are widely used to parameterize convection in coarse resolution ocean models. Results are not found to be sensitive to this choice as long as the associated diffusive time scale $D^2/\nu_{\mathrm{conv}}\approx 0.5$~yr is much shorter than the advective time scale $M_{\mathrm{half}}/\Psi\approx 2000$~yrs ($M_{\mathrm{half}}$ is half of the total mass of the ocean and $\Psi$ is the maximum meridional streamfunction in $kg/s$).


  The horizontal and vertical viscosity are set to $1$~m$^2$/s to retain a smooth solution on the grid-scale and to make sure that the Ekman boundary layer is thick enough to be resolved. To damp grid-size noise induced by stair-like topography, we introduce a bi-harmonic hyperviscosity of $1\times 10^8$~m$^4$/s and a bi-harmonic hyperdiffusivity of $1\times 10^7$~m$^4$/s in addition to the harmonic mixing terms. Adding a scale-selective viscosity stabilizes the model and smooths the solution, whilst retaining large-scale features. 
  
\begin{figure}
  \center
\includegraphics[width=0.8\textwidth,page=7]{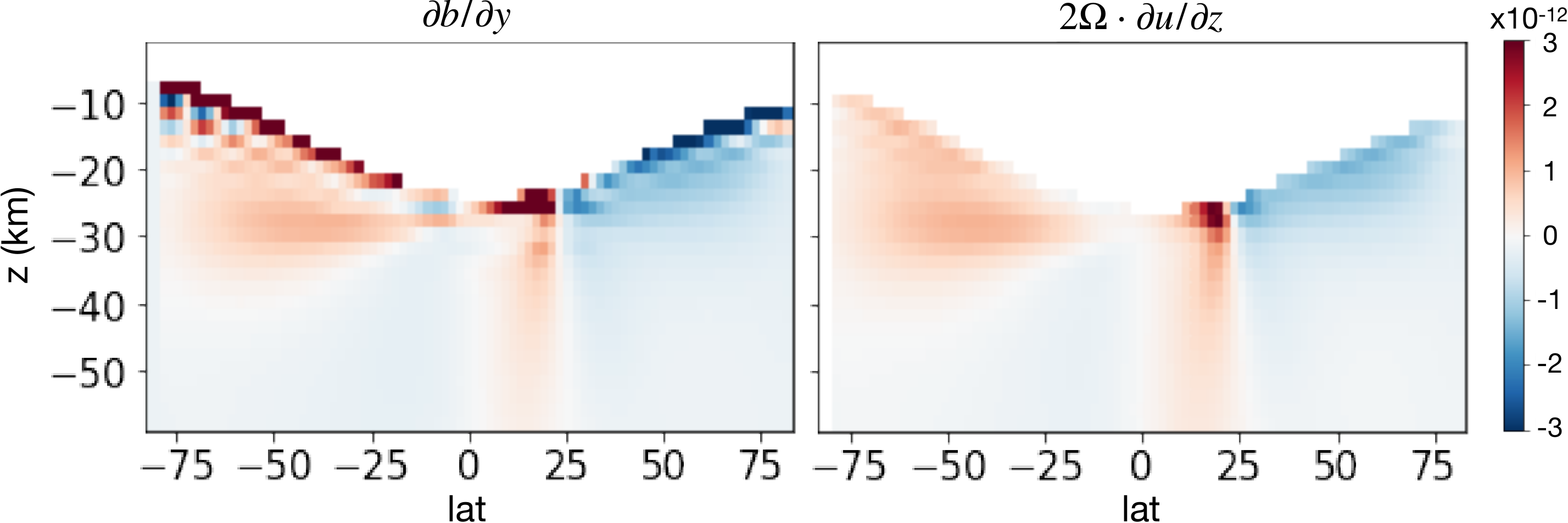}
\caption{Balance of terms in the thermal wind balance. Left and right panels show the two terms in the thermal wind balance, $2\mathbf{\Omega}\cdot \nabla U$ and $\partial b/a\partial \theta$, respectively. $\Omega$ is the rotating angular speed, $U$ is the zonal flow speed, $b=-g(\rho-\rho_0)/\rho_0$ is buoyancy, $a$ is the moon's radius and $\theta$ denotes latitude. Away from the very near surface layers, there is very good balance.}
\label{fig:thermal-wind}
\end{figure}

  To demonstrate that the viscous terms indeed play a minor role, Fig.~\ref{fig:thermal-wind} shows, for one example case (S30s80), that the two terms in the thermal wind balance --- $2\mathbf{\Omega}\cdot \nabla U=\partial b/a\partial \theta$ --- balance one-another. Since thermal wind is a consequence of geostrophic and hydrostatic balance, and on the large scale hydrostatic balance is always a good approximation, geostrophic balance must also be satisfied.

  \subsection{Heat budget}
  \label{sec:ice-heat-budget}
  Heat is lost in our system through heat conduction across the ice shell, $\mathcal{H}_{\mathrm{cond}}$, which can be estimated using a 1D vertical heat conduction model,
\begin{equation}
  \mathcal{H}_{\mathrm{cond}}=\frac{\kappa_{0}}{H} \ln \left(\frac{T_{f}}{T_{s}}\right),
  \label{eq:H-cond}
  \end{equation}
  given the thickness of the ice $H$ \citep{Hemingway-Mittal-2019:enceladuss} (black solid curve in Fig.~\ref{fig:EOS-Hice-Heatflux}b), the surface $T_s$ (set to the radiative equilibrium temperature of Enceladus assuming an albedo of 0.8) and an ice temperature near the water-ice interface, which, by definition, is the local freezing point $T_f$ given by Eq.~\ref{eq:freezing-point}. This yields a heat loss rate of roughly $50$~mW/m$^2$, as shown by the green dashed curve in Fig.~\ref{fig:EOS-Hice-Heatflux}b of the main text).
  
This loss is counterbalanced by tidal heating in the ice shell, $\mathcal{H}_{\mathrm{ice}}$. and the core, $\mathcal{H}_{\mathrm{core}}$. Heating in the shell is forced by ocean tides \footnote{There have been some misunderstanding in the literature regarding whether crust (i.e., the ice shell) dissipation driven by ocean tide should be counted as ice dissipation or ocean dissipation. A detailed literature review and discussion is available in \cite{Tyler-2020:heating}.}, but tidal dissipation within the ocean itself is likely to play only a secondary role due to the presence of the the ice shell \citep{Chen-Nimmo-2011:obliquity, Beuthe-2016:crustal, Hay-Matsuyama-2019:nonlinear, Rekier-Trinh-Triana-et-al-2019:internal}, except if the ocean is in a resonant state \cite{Tyler-2008:strong, Tyler-2011:tidal, Tyler-2014:comparative}. For each assumed heat partition, we use the same heating profiles for $\mathcal{H}_{\mathrm{core}}$ and $\mathcal{H}_{\mathrm{ice}}$ (see below).

According to \citet{Beuthe-2019:enceladuss} and \citet{Choblet-Tobie-Sotin-et-al-2017:powering}, the core dissipation $\mathcal{H}_{\mathrm{core}}$ peaks at the two poles. We obtain the meridional heat profile using Eq.60 in \citet{Beuthe-2019:enceladuss} (Beuthe, 2020, private communication,
  \begin{equation}
    \label{eq:H-core}
    \mathcal{H}_{\mathrm{core}}(\phi)= \mathcal{H}_{\mathrm{core}}(\phi)=\overline{\mathcal{H}_{\mathrm{core}}}\cdot (1.08449 + 0.252257 \cos(2\phi) + 0.00599489 \cos(4\phi)),
  \end{equation}
where $\phi$ is the latitude and $\overline{\mathcal{H}_{\mathrm{core}}}$ is the global mean heat flux at the bottom. Since the global surface area shrinks going downward due to the spherical geometry, a factor of $\left.(a-H)^2\right/(a-H-D)^2$ ($H$ is ice thickness, $D$ is ocean depth) needs to be taken into account when computing $\overline{\mathcal{H}_{\mathrm{core}}}$. The expression within the brackets above is normalized over the globe, and adjusted to take account of the fact that our model only covers 84S-84N. The above formula yields a bottom heat flux that is polar-amplified, as shown by the red dashed curve in Fig.~\ref{fig:EOS-Hice-Heatflux}b of the main text. This is prescribed as an upward geothermal heat flux from the lower boundary.

The remaining heat loss is balanced by tidal dissipation in the ice shell $\mathcal{H}_{\mathrm{ice}}$. We calculate $\mathcal{H}_{\mathrm{ice}}$ using the thin ice shell model described in \citet{Beuthe-2019:enceladuss}.
Tidal dissipation consists of three components \citep{Beuthe-2019:enceladuss}: a membrane mode $\mathcal{H}_{\mathrm{ice}}^{\mathrm{mem}}$ due to the extension/compression and tangential shearing of the ice membrane, a mixed mode $\mathcal{H}_{\mathrm{ice}}^{mix}$ due to vertical shifting, and a bending mode $\mathcal{H}_{\mathrm{ice}}^{bend}$ induced by the vertical variation of compression/stretching. Following \citet{Beuthe-2019:enceladuss}, we first assume the ice sheet to be completely flat and solve for the dissipation rate $\mathcal{H}_{\mathrm{ice}}^{\mathrm{flat,x}}$ (where $x=\{\mathrm{mem},\mathrm{mix},\mathrm{bend}\}$). The ice properties are derived assuming a globally-uniform surface temperature of 60K and a melting viscosity of $5\times10^{13}$~Pa$\cdot$s. 

Ice thickness variations are accounted for by multiplying the membrane mode dissipation $\mathcal{H}_{\mathrm{ice}}^{\mathrm{flat,mem}}$, by a factor that depends on ice thickness. This makes sense because it is the only mode which is amplified in thin ice regions \citep[see ][]{Beuthe-2019:enceladuss}. This results in the expression:
\begin{equation}
  \label{eq:H-tide}
  \mathcal{H}_{\mathrm{ice}}=(H/H_0)^{p_\alpha}\mathcal{H}_{\mathrm{ice}}^{\mathrm{flat,mem}}+\mathcal{H}_{\mathrm{ice}}^{\mathrm{flat,mix}}+\mathcal{H}_{\mathrm{ice}}^{\mathrm{flat,bend}},
\end{equation}
where $H$ is the prescribed thickness of the ice shell as a function of latitude and $H_0$ is the global mean of $H$. The tidal heating profiles are given as follows,
\begin{eqnarray}
  \mathcal{H}_{\mathrm{ice}}^{\mathrm{flat,mem}}&=&A\left(Y_{00}+0.25Y_{20}+0.0825Y_{40}\right)\label{eq:H-tide-mem}\\
  \mathcal{H}_{\mathrm{ice}}^{\mathrm{flat,mem}}+\mathcal{H}_{\mathrm{ice}}^{\mathrm{flat,bend}}&=&A\left(0.124Y_{00}+0.196Y_{20}-0.0199Y_{40}\right),\label{eq:H-tide-mixbend}
\end{eqnarray}
where $Y_{00},\ Y_{20},\ Y_{40}$ are spherical harmonics and $A$ is a constant factor used to adjust the amplitude of $\mathcal{H}_{\mathrm{ice}}$.

Since thin ice regions deform more easily and produce more heat, $p_\alpha$ is negative. Because more heat is produced in the ice shell, the overall ice temperature rises, which, in turn, further increases the mobility of the ice and leads to more heat production (the rheology feedback). We set $p_\alpha=-2$ here. $\mathcal{H}_{\mathrm{ice}}$, rescaled to balance $\mathcal{H}_{\mathrm{cond}}$, is shown by the blue solid curve in Fig.~\ref{fig:EOS-Hice-Heatflux}b. Those who are interested in further details of the model are encouraged to read \citet{Beuthe-2019:enceladuss}.

  \subsection{Water-ice interaction}

  The interaction between the ocean and ice is simulated using MITgcm's ``shelf-ice'' package \citep{Losch-2008:modeling, Holland-Jenkins-1999:modeling}. We turn on the ``boundary layer'' option to avoid possible numerical instabilities induced by an ocean layer which is too thin. The code is modified to account for a gravitational acceleration that is very different from that on earth, the temperature dependence of heat conductivity, and the meridional variation of tidal heating generated inside the ice shell and the ice surface temperature. Following \citet{Kang-Bire-Campin-et-al-2020:differing}, the freezing/melting rate of the ice shell is determined by a heat budget for a thin layer of ice at the base\footnote{This choice is supported by the fact that most tidal heating is generated close to the ocean-ice interface \citep{Beuthe-2018:enceladuss}.}. The budget involves three terms: the heat transmitted from the ocean to the ice $\mathcal{H}_{\mathrm{ocn}}$ (positive upward), the heat loss through the ice shell due to heat conduction $\mathcal{H}_{\mathrm{cond}}$ (Eq.\ref{eq:H-cond}), and the tidal heating generated inside the ice shell $\mathcal{H}_{\mathrm{ice}}$ (Eq.\ref{eq:H-tide}). As elucidated in \citet{Holland-Jenkins-1999:modeling} and \citet{Losch-2008:modeling}, the continuity of heat flux and salt flux through the ``boundary layer'' gives,
  \begin{eqnarray}
    &~&\mathcal{H}_{\mathrm{ocn}}-\mathcal{H}_{\mathrm{cond}}+\mathcal{H}_{\mathrm{ice}}=-L_fq-C_p(T_{\mathrm{ocn-top}}-T_b)q\label{eq:boundary-heat}\\
&~&\mathcal{F}_{\mathrm{ocn}}=-S_bq-(S_{\mathrm{ocn-top}}-S_b)q, \label{eq:boundary-salinity}   
  \end{eqnarray}
  where $q$ denotes the freezing rate in $kg/m^2/s$, $\mathcal{F}_{\mathrm{ocn}}$ denotes the salinity flux (positive upward), $T_{\mathrm{ocn-top}}$ and $S_{\mathrm{ocn-top}}$ is the temperature and salinity in the top grid of the ocean\footnote{When model resolution is smaller than the boundary layer thickness, the salinity below the upper-most grid cell also contributes to $T_{\mathrm{ocn-top}}$ and $S_{\mathrm{ocn-top}}$.}, $T_b$ and $S_b$ denote the temperature and salinity in the imaginary ``boundary layer'', where the water is just at the freezing point. $C_p=4000$~J/kg/K is the heat capacity of the ocean, $L_f=334000$~J/kg is the latent heat of fusion of ice. The last terms in Eq.\ref{eq:boundary-heat} and Eq.\ref{eq:boundary-salinity} are associated with the warming/cooling and salinification/freshening required to convert ocean water to water in the ``boundary layer''.
  
$\mathcal{H}_{\mathrm{ocn}}$ and $\mathcal{F}_{\mathrm{ocn}}$ in Eq.\ref{eq:boundary-heat} can be written as
  \begin{eqnarray}
    \mathcal{H}_{\mathrm{ocn}}&=&C_p(\rho_{0}\gamma_T-q)(T_{\mathrm{ocn-top}}-T_b),\label{eq:H-ocn}\\
    \mathcal{F}_{\mathrm{ocn}}&=&(\rho_{0}\gamma_S-q)(S_{\mathrm{ocn-top}}-S_b) \label{eq:S-ocn}
  \end{eqnarray}
  where $\gamma_T=\gamma_S=10^{-5}$~m/s are the exchange coefficients for temperature and salinity, and $T_b$ denotes the and temperature in the ``boundary layer''. The terms associated with $q$ are the heat/salinity change induced by the deviation of $T_{\mathrm{ocn-top}},\ S_{\mathrm{ocn-top}}$ from that in the ``boundary layer'', where melting and freezing occur. $T_b$ equals the freezing temperature $T_f(S_b,P)$ (Eq.\ref{eq:freezing-point}) at pressure $P$ and salinity $S_b$ by definition. $\rho_0$ is the reference density (Eq.\ref{eq:rho0}).

  The only two unknowns, $S_b$ and $q$, in Eq.~(\ref{eq:boundary-heat}) and Eq.~(\ref{eq:boundary-salinity}) can therefore be solved jointly. When freezing occurs ($q>0$), salinity flux $\rho_{w0}\gamma_S(S_{\mathrm{ocn-top}}-S_b)$ is negative (downward). This leads to a positive tendency of salinity at the top of the model ocean, together with changes of temperature.
  \begin{eqnarray}
    \frac{dS_{\mathrm{ocn-top}}}{dt}&=&\frac{-\mathcal{F}_{\mathrm{ocn}}}{\rho_{w0}\delta z}=\frac{1}{\rho_{w0}\delta z}(\rho_{w0}\gamma_S-q)(S_b-S_{\mathrm{ocn-top}})=\frac{qS_{\mathrm{ocn-top}}}{\rho_{w0}\delta z},\label{eq:S-tendency}\\
    \frac{dT_{\mathrm{ocn-top}}}{dt}&=&\frac{-\mathcal{H}_{\mathrm{ocn}}}{C_p\rho_{w0}\delta z}=\frac{1}{\rho_{w0}\delta z}(\rho_{w0}\gamma_T-q)(T_b-T_{\mathrm{ocn-top}})\nonumber\\
    &=&\frac{1}{C_p\rho_{w0}\delta z}\left[\mathcal{H}_{\mathrm{ice}}-\mathcal{H}_{\mathrm{cond}}+L_fq+C_p(T_{\mathrm{ocn-top}}-T_b)q\right]
  \end{eqnarray}
  where $\delta z$ is the thickness of the upmost ocean grid.

The flow speed is relaxed to zero at the top boundary at a rate of $\gamma_M=1\times10^{-3}$~m/s following
\begin{equation}
    \frac{\partial u}{\partial t}=\ldots -\frac{\gamma_M}{\delta z} u.
\end{equation}

\begin{table}[hptb!]
  
  \centering
  \begin{tabular}{lll}
    \hline
    Symbol & Name & Definition/Value\\
    \hline
    $a$ & radius of Enceladus & 252~km\\
    $\delta$ & obliquity of Enceladus & 27$^\circ$\\
    $A$ & albedo of Enceladus & 0.81 \\ 
    $H$ & global averaged ice shell thickness & 20.8~km $^*$  \\
    $D$ & global averaged ocean depth & 39.2~km $^*$  \\
    $\Omega$ & rotation rate & 5.307$\times$10$^{-5}$~s$^{-1}$\\
    $g_0$ & surface gravity & 0.113~m/s$^2$\\
    $g$ & gravity in the ocean & Eq.\ref{eq:g-z}\\
    $L_f$ & latent heat of fusion of ice & 334000~J/kg\\
    $C_p$ & heat capacity of water & 4000~J/kg/K\\
    $T_f(S,P)$ & freezing point & Eq.\ref{eq:freezing-point}\\
    $\rho_i$ & density of ice & 917~kg/m$^3$ \\
    $\rho_w$ & density of the ocean & Eq.\ref{eq:EOS-linear}\\
    $S_0$ & mean ocean salinity & 5, 10, 15, 20, 25, 30~g/kg (psu) \\
    $P_0$ & reference pressure & $\rho_ig_0H=2.16\times10^6$~Pa \\
    $T_0$ & reference temperature & $T_f(S_0,P_0)$ \\
    $\rho_{w0}$ & reference density of ocean & Eq.\ref{eq:rho0} \\
    $\alpha,\beta$ & thermal expansion \& saline contraction coeff. &  using Gibbs Seawater Toolbox $^{**}$  \\
    $\kappa_0$ & conductivity coeff. of ice & 651~W/m $^{***}$\\
    $\nu_h,\ \nu_v$ & horizontal/vertical viscosity & 50~m$^2$/s\\
    $\tilde{\nu}_h,\ \tilde{\nu}_v$ & bi-harmonic hyperviscosity & 3$\times$10$^9$~m$^4$/s\\ 
    $\kappa_h,\ \kappa_v$ & horizontal/vertical diffusivity & 0.001~m$^2$/s\\
    $(\gamma_T,\ \gamma_S,\ \gamma_M)$ & water-ice exchange coeff. for T, S \& momentum & (10$^{-5}$, 10$^{-5}$, 10$^{-3}$)~m/s\\
    $\mathcal{H}_{\mathrm{cond}}$ & conductive heat loss through ice & Eq.\ref{eq:H-cond}\\
    $\mathcal{H}_{\mathrm{ice}}$ & tidal heating produced in the ice & Eq.\ref{eq:H-tide} \\
    $\mathcal{H}_{\mathrm{core}}$ & bottom heat flux powered by the core & Eq.\ref{eq:H-core} \\
    $p_\alpha$ & ice dissipation amplification factor & -2  \\
    \hline
     \end{tabular}
  \caption{Default model parameters. Ref $^*$: \citet{Hemingway-Mittal-2019:enceladuss}, ref $^{**}$: \citet{McDougall-Barker-2011:getting}, ref $^{***}$: \citet{Petrenko-Whitworth-1999:physics}.}
  \label{tab:parameters}
\end{table}

\section{Description of the ice flow model.}
 To compute the down-slope mass transport induced by ice thickness variation, we use an upside-down land ice sheet model following \citet{Ashkenazy-Sayag-Tziperman-2018:dynamics}. The ice flows down its thickness gradient, driven by the pressure gradient induced by the spatial variation of the ice top surface, somewhat like a second order diffusive process. At the top, the speed of the ice flow is negligible because the upper part of the shell is so cold and hence rigid; at the bottom, the vertical shear of the ice flow speed vanishes, as required by the assumption of zero tangential stress there. This is the opposite to that assumed in the land ice sheet model. In rough outline, we calculate the ice flow using the expression below obtained through repeated vertical integration of the force balance equation (the primary balance is between the vertical flow shear and the pressure gradient force), using the aforementioned boundary conditions to arrive at the following formula for ice transport $\mathcal{Q}$,
\begin{eqnarray}
  \mathcal{Q}(\phi)&=& \mathcal{Q}_0H^3(\partial_\phi H/a) \label{eq:ice-flow}\\
  \mathcal{Q}_0&=&\frac{2(\rho_0-\rho_i)g}{\eta_{\mathrm{melt}}(\rho_0/\rho_i)\log^3\left(T_f/T_s\right)}\int_{T_s}^{T_f}\int_{T_s}^{T(z)}\exp\left[-\frac{E_{a}}{R_{g} T_{f}}\left(\frac{T_{f}}{T'}-1\right)\right]\log(T')~\frac{dT'}{T'}~\frac{dT}{T}.\nonumber 
\end{eqnarray}
Here, $\phi$ denotes latitude, $a=252$~km and $g=0.113$~m/s$^2$ are the radius and surface gravity of Enceladus, $T_s$ and $T_f$ are the temperature at the ice surface and the water-ice interface (equal to local freezing point, Eq.~\ref{eq:freezing-point}), and $\rho_i=917$~kg/m$^3$ and $\rho_0$ are the ice density and the reference water density (Eq.~\ref{eq:EOS-linear}). $E_a=59.4$~kJ/mol is the activation energy for diffusion creep, $R_g=8.31$~J/K/mol is the gas constant and $\eta_{\mathrm{melt}}$ is the ice viscosity at the freezing point. The latter has considerable uncertainty \citep[$10^{13}$-$10^{16}$~Pa$\cdot$s][]{Tobie-Choblet-Sotin-2003:tidally} but we choose to set $\eta_{\mathrm{melt}}=10^{14}$~Pa$\cdot$s.

\newpage
\bibliographystyle{cas-model2-names}
\bibliography{export}
\end{document}